\documentclass[aps,prl,twocolumn,amsmath,amssymb,superscriptaddress,floatfix]{revtex4}
\usepackage{graphicx}
\usepackage{amssymb}
\usepackage{natbib}
\usepackage{color}
\usepackage{float}
\usepackage{placeins}

\begin{document}

\title{Incommensurate Spin Fluctuations in the Spin-triplet Superconductor Candidate UTe$_2$}

\author{Chunruo Duan}
\affiliation{Department of Physics and Astronomy, Rice University, Houston, TX 77005, USA}
\author{Kalyan Sasmal}
\author{M. Brian Maple}
\affiliation{Department of Physics, University of California, San Diego, CA 92093, USA}
\author{Andrey Podlesnyak}
\affiliation{ Neutron Scattering Sciences Division, Oak Ridge National Laboratory, Oak Ridge, TN 37831, USA}
\author{Jian-Xin Zhu}
\affiliation{Theoretical Division and Center for Integrated Nanotechnologies, Los Alamos National Laboratory, Los Alamos, NM 87545, USA}
\author{Qimiao Si}
\author{Pengcheng Dai}
\email{pdai@rice.edu}
\affiliation{Department of Physics and Astronomy, Rice University, Houston, TX 77005, USA}

\date{\today}

\begin{abstract}
Spin-triplet superconductors are of extensive current interest because 
they can host topological
state and Majorana ferimons important for quantum computation. 
The uranium based heavy-fermion superconductor UTe$_2$ has been argued as a 
spin-triplet superconductor similar to UGe$_2$, URhGe, and UCoGe, 
where the superconducting phase is near (or coexists with) a ferromagnetic (FM) 
instability and spin-triplet electron pairing is 
driven by FM spin fluctuations. Here we use neutron scattering to show that 
although UTe$_2$ exhibits no static magnetic order down to 0.3 K, its magnetism is dominated by incommensurate spin fluctuations near antiferromagnetic (AF) ordering wave vector and  
extends to at least 2.6 meV. We are able to understand the dominant incommensurate spin fluctuations of UTe$_2$ in terms of its electronic structure calculated using 
a combined density functional and dynamic mean field theory.
\end{abstract}

\maketitle

Superconductivity occurs in many metals when electrons form
coherent Cooper pairs below the superconducting 
transition temperature $T_c$ \cite{BCS}. In conventional and most unconventional superconductors, electron Cooper pairs in the superconducting state 
form anti-parallel spin-singlets with the total spin $S=0$ \cite{scalapinormp,Lee-rmp,Si-nrm,Stewart,Dai2015}. However, electrons in the superconducting state can also form parallel spin-triplet Cooper pairs \cite{Fay1980,andy,Kallin}, analogous to the equal spin pairing state in the superfluid $^3$He \cite{Lee1997}. The Pauli's exclusion principle
can be fulfilled for both singlet and triplet Cooper pairs 
by adjusting the symmetry of the orbital part of the wave function. For the 
spin-singlet 
pairing state, the orbital wave function
has even parity (symmetric) with orbital 
angular momentum $L=0$ ($s$-wave), 2 ($d$-wave), etc. For spin-triplet state, the orbital wave function has odd parity
 (antisymmetric) with orbital angular momentum $L=1$ ($p$-wave), 3 ($f$-wave), and etc \cite{andy,Kallin}. While most unconventional superconductors have
spin-singlet pairing associated with antiferromagnetic (AF) 
spin fluctuations \cite{scalapinormp,Lee-rmp,Si-nrm,Stewart,Dai2015}, spin-triplet superconductors are rare, and the superconductivity is believed to be driven by longitudinal ferromagnetic (FM) spin fluctuations \cite{Fay1980,andy,Kallin}. Since spin-triplet superconductors are intrinsically
topological \cite{Fu2010,MSato2010,XLQi2010,MSato2017} and can host Majorana fermions
important for quantum computation \cite{NRead2000,Kitaev2001}, it is important
to understand a candidate spin-triplet superconductor by determining the associated spin fluctuations.

In spin-triplet superconductors such as UGe$_2$ \cite{Saxena2000}, 
URhGe \cite{Aoki2001}, and UCoGe \cite{Huy2007}, superconductivity arises through
suppression of the static FM order \cite{Saxena2000} or coexists 
with static FM order \cite{Aoki2001,Huy2007}. Inelastic neutron 
scattering (INS) experiments find 
clear evidence of FM spin fluctuations in URhGe \cite{Huxley2003} and UCoGe \cite{Stock2011}. For the spin-triplet candidate superconductor Sr$_2$RuO$_4$ \cite{andy}, 
where the material is paramagnetic at all temperatures and 
superconductivity does not coexist with static FM order,
magnetism is dominated by incommensurate spin fluctuations
arising from Fermi surface 
nesting of itinerant electrons \cite{Braden2004,Kunkem2017,Iida2020}, although weak FM spin fluctuations are also observed \cite{Steffens2019}. 
Similarly, although considerable evidence exists for spin-triplet superconductivity in
UPt$_3$ \cite{Joynt2002,ERSchemm2014,Avers2020}, its superconductivity appears to be associated with AF order and spin fluctuations 
instead of FM spin fluctuations \cite{Aeppli1988,Aeppli1989}.

Recently, UTe$_2$ has been identified as a new spin-triplet superconductor
with $T_c\approx 1.6$ K \cite{Ran2019,Aoki2019,Ran2019b,Imajo2019}. 
UTe$_2$ has an orthorhombic unit cell with space group $Immm$,
where the U atoms form parallel ladders along the $a$-axis inside 
trigonal prisms of Te atoms [Fig. 1(a)] \cite{Ikeda2006}.
 The shortest U-U bond is along the rung of the ladder in the $c$-axis direction, while the easy-axis of the U spins is along the $a$-axis. 
The symmetry operation that connects one ladder to its nearest neighbor is the body-center $(\frac12,\frac12,\frac12)$ translation. 
A Curie-Weiss fit to the magnetic susceptibility data reveals an effective moment per U atom close to the $5f^2$ or $5f^3$ free ion value at high temperature \cite{Aoki2019}. No long-range magnetic order has been reported down to 0.25 K 
\cite{Ran2019,Aoki2019,Ran2019b,Imajo2019,Ikeda2006,Sundar2019}. Instead, a 
sudden increase in the magnetic susceptibility below 10 K 
in response to a magnetic field applied parallel to the $a$-axis resembles the quantum critical behavior of metallic ferromagnets, indicating strong FM spin fluctuations along the $a$-axis \cite{Ran2019}. This suggests that UTe$_2$ sits at the paramagnetic end of a series of FM heavy fermion superconductors including UGe$_2$ \cite{Saxena2000}, URhGe \cite{Aoki2001}, 
and UCoGe \cite{Huy2007}. At the FM end, the compound UGe$_2$ is a pressure-induced superconductor with optimal $T_c\approx 0.5$ K at 1.2 GPa \cite{Tateiwa_2000,Bauer2001}. 
Moving from UGe$_2$ to URhGe, superconductivity occurs at ambient pressure 
below $T_c\approx 0.25$ K and coexists with static FM order below a Curie temperature $T_C\approx 9.5$ K \cite{Aoki2001}. Finally, UCoGe has coexisting superconductivity and
FM order with increased $T_c\approx 0.425$ K and decreased $T_C\approx 3$ K, respectively \cite{Huy2007}.

The scenario that UTe$_2$ is a candidate
spin-triplet superconductor \cite{Ran2019,Aoki2019} is supported 
by a growing list of observations. These include the 
upper critical fields $H_{C2}$ that exceed the Pauli limits 
along all crystallographic directions \cite{Ran2019b,Imajo2019};
temperature independent $^{125}$Te Knight shift across $T_c$ 
in nuclear magnetic resonance measurements  \cite{Ran2019}; 
coexisting FM spin fluctuations and superconductivity \cite{Sundar2019,Tokunaga2019}; 
signatures of chiral superconductivity \cite{LJiao2020}; and breaking of time reversal symmetry expected for a spin-triplet superconductor \cite{Hayes2020}. There are also  theoretical \cite{yang2019} and experimental \cite{wray2020} efforts 
to understand the underlying electronic structure of UTe$_2$.

In this Letter, we use INS to probe the wave vector and energy
dependence of spin fluctuations in UTe$_2$. In addition to confirming 
that UTe$_2$ exhibits no static magnetic order down to 0.3 K, 
we discovered that the dominant spin fluctuations in UTe$_2$ are three-dimensional (3D)
in reciprocal space, centered at the 
incommensurate wave vector ${\bf Q}=(0,\pm(K+ 0.57),0)$ ($K=0,1$),
and extend to energies of at least $E= 2.6$ meV. 
FM spin fluctuations, if present, are much weaker than the incommensurate spin fluctuations. 
Based on density functional theory (DFT) in the
generalized gradient approximation (GGA)~\cite{JPPerdew:1996}, combined with dynamical 
mean-field theory (DMFT) calculations \cite{AGeorges:1996,GKotliar:2006,KHaule:2010,RTutchton:2020}, we
understand the dominant incommensurate 
spin fluctuations by showing that the associated wave vector is approximately consistent 
with the AF wave vector of the Ruderman-Kittel-Kasuya-Yosida (RKKY)
 interaction between the $5f$ moments.
Therefore, in addition to a FM instability, 
incommensurate (close to AF) spin fluctuations must also be considered to unveil the
magnetic and superconducting properties of UTe$_2$.  

\begin{figure}[t]
\includegraphics[width=0.95\linewidth]{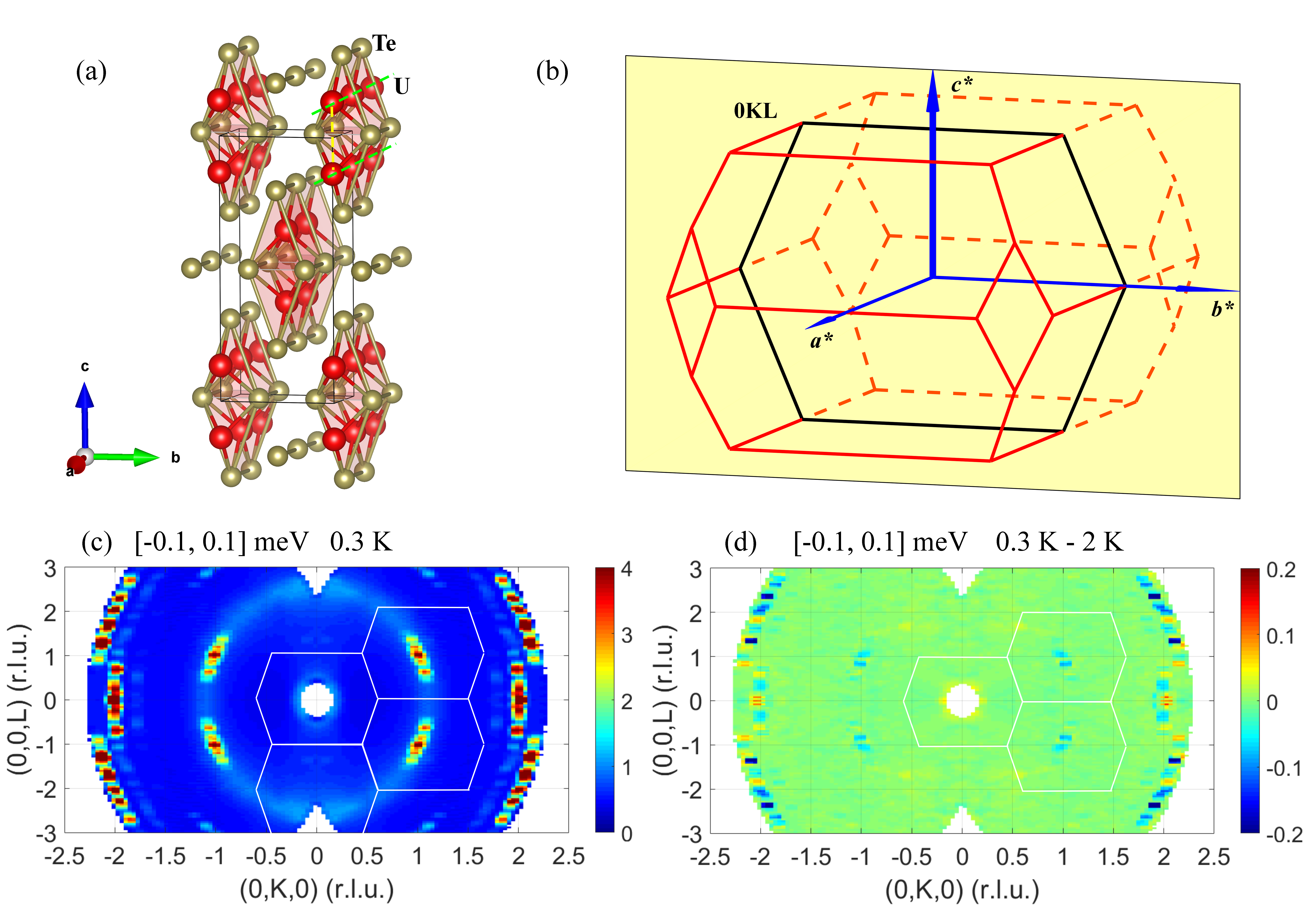}
\caption{(a) The crystal structure of UTe$_2$. The first Brillouin zones (BZ) is sketched in (b), with the edges behind the $(0,K,L)$ plane plotted as dashed lines. The reciprocal
space is labeled as $\mathbf{a}^*$, $\mathbf{b}^*$, $\mathbf{c}^*$. (c) The diffraction pattern of UTe$_2$ in the $(0,K,L)$ plane at $T=0.3$ K. The integration range along the $H$ direction is from -0.1 to 0.1 r.l.u., and along energy is from -0.1 to 0.1 meV. (d) The temperature difference spectra between 
0.3 K and 2 K. The BZs are indicated by white solid lines in (c) and (d). }
\end{figure}

Our INS experiments were carried out at the Cold Neutron Chopper Spectrometer (CNCS) at Oak Ridge National Laboratory. 
The momentum transfer ${\bf Q}$ in 3D reciprocal 
space is defined as
${\bf Q}$ = $H{\bf a}^* + K{\bf b}^* + L{\bf c}^*$, where $H, K$ and $L$ are Miller indices and ${\bf a}^* = \hat {\bf a}2\pi/a$, ${\bf b}^* = \hat {\bf b}2\pi/b$, ${\bf c}^* = \hat {\bf c}2\pi/c$ with $a$ = 4.16 \AA, $b$ = 6.12 \AA\ and $c$ = 13.95 \AA\ of the orthorhombic lattice \cite{Ikeda2006}. 
Single crystals of UTe$_2$ were prepared using the chemical vapor transport method with I$_2$ as the transport media \cite{SI}.
The crystals are naturally cleaved along the $ab$-plane and form small flakes of about 0.5 to 1 mm thick. The typical dimension of the crystals in the $ab$-plane is from 1 to 2 mm and the mass is in the range of 10 to 30 mg. We co-aligned 61 pieces (total mass 0.7-g) of single crystals on oxygen-free Cu-plates using an X-ray Laue machine to check
the orientation of each single crystal. The sample assembly is aligned in the
$[0,K,L]$ scattering plane as shown in the black frame of Fig. 1(b) and mounted inside a He$^3$ refrigerator. 
Most of our measurements were carried out with $E_i=3.37$ meV on CNCS at different temperatures.

\begin{figure}[t]
\includegraphics[width=0.95\linewidth]{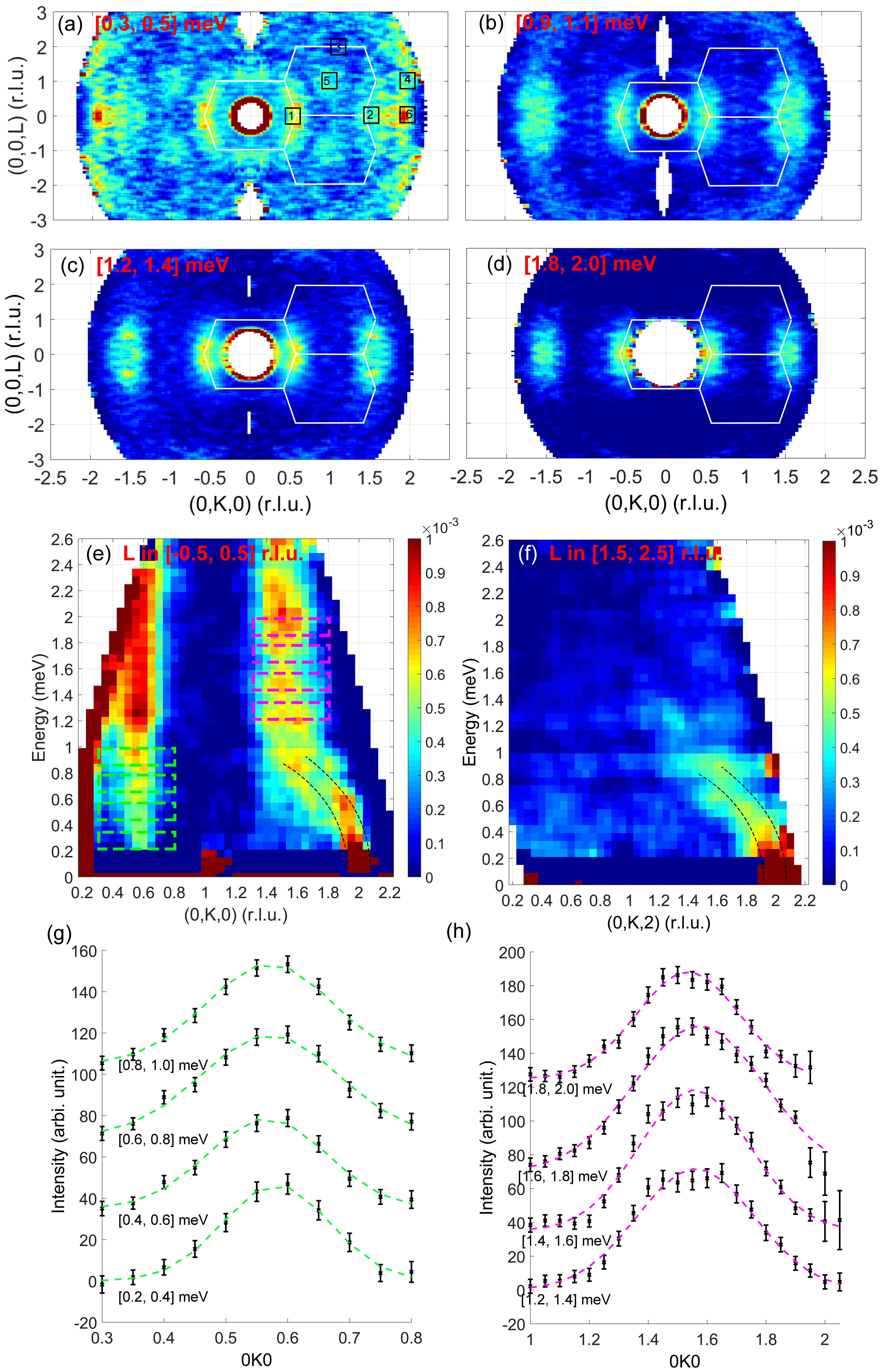}
\caption{Images of 2D constant-energy cuts in the $(0,K,L)$ scattering plane at $T=0.3$ K. Along the $H$ direction, the integration range is from -0.1 to 0.1 $r.l.u.$, while along energy, the ranges are (a) 0.3 to 0.5 meV, (b) 0.9 to 1.1 meV, (c) 1.2 to 1.4 meV, and (d) 1.8 to 2.0 meV. The BZs are indicated by white solid lines. In subplot (a), black squares labeled from 1 to 6 indicate the ${\bf Q}$ ranges of the 1D cuts along energy 
plotted in Fig. 3(a-f), respectively. $\chi^{\prime\prime}(E,{\bf Q})$ integrated in (e) $-0.5 < L < 0.5$ and (f) $1.5 < L < 2.5$ r.l.u.  (g,h) 1D cuts along the $K$ direction at different $E$ and ${\bf Q}$ across ${\bf Q}_{\rm IC}$. Their corresponding integration ranges are marked by dashed rectangles of the same color in (e). Backgrounds fitted by linear functions are subtracted in the 1D cuts, and the curves are artificially separated 
along the $y$-axis for clarity. By fitting the peaks with Gaussian functions, we obtain $K=0.57\pm0.01$ and $1.56\pm0.01$ r.l.u.}
\end{figure}

Figure 1(c) shows a map of reciprocal space in the $[0,K,L]$ scattering plane 
at 0.3 K and elastic position.
We see nuclear Bragg peaks at 
the expected $(0,\pm 1,\pm 1)$, $(0,\pm 2,0)$, and $(0,\pm2,\pm2)$ positions.
The spread of the Bragg peaks along the $L$ direction indicates 
a broad sample mosaic of $\sim$15 degrees. 
To search for possible static FM or AF magnetic order, we show in Fig. 1(d) the temperature
difference plot between 0.3 K and 2 K, and find no evidence of intensity gain anywhere 
within the probed reciprocal space. We therefore conclude that UTe$_2$ does not exhibit
static FM or AF order down to 0.3 K, consistent with earlier work \cite{Ran2019,Aoki2019,Ran2019b,Imajo2019,Sundar2019}.

\begin{figure}[t]
\includegraphics[width=0.95\linewidth]{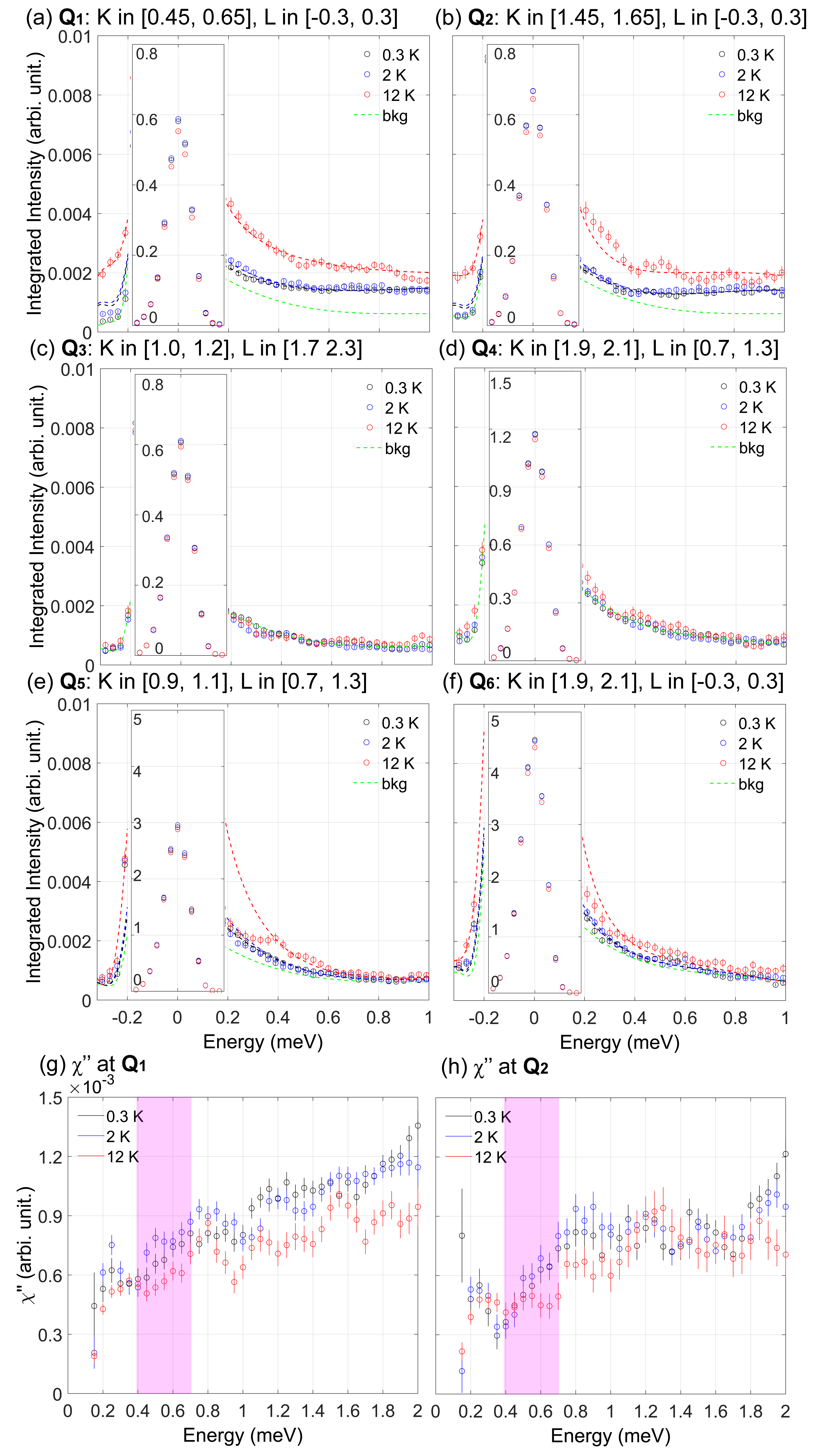}
\caption{(a-f) Constant-{\bf Q} cuts correspond to the ${\bf Q}_1$-${\bf Q}_6$ positions
marked in Fig. 2(a) at $T=0.3, 2$, and 12 K. The insets shows the incoherent or Bragg intensity from $E=-0.2$ to 0.2 meV depending on whether a Bragg peak exists at each ${\bf Q}$ position. Dashed lines on the neutron energy loss side ($E>0$ side) are obtained by fitting the $T=0.3$ K data, which are then scaled up based on the Bose population factor for the 2 K and 12 K data. On the neutron energy gain side ($E<0$), dashed lines are obtained by fitting the 12 K data, which are then scaled down based on the Bose population factor. (g,h) The $\chi^{\prime\prime}(E)$ within $\mathbf{Q}_1$ and $\mathbf{Q}_2$, respectively. The pink shadow region highlights the energy range of $3k_BT_c$ to $5k_BT_c$, where a spin resonance appears below $T_c$ in many 
spin-singlet unconventional superconductors \cite{Eschrig2006,Wilson2006,GYu2009,Inosov2011}.}
\end{figure}

Despite the absence of long range magnetic order, INS experiments on UTe$_2$
reveal clear evidence for excitations 
at finite energy transfers. Figures 2(a-d) show 2D images of  
constant-energy cuts in the $[0,K,L]$ plane at different 
energies below $T_c$ ($T=0.3$ K).
At $E=0.4\pm 0.1$ meV, we see clear excitations at incommensurate 
wave vectors ${\bf Q}_{\rm IC}=(0,\pm 0.57,0)$ and a possible signal at FM wave vectors 
 ${\bf Q}_{\rm FM}=(0,\pm 1,\pm 1)$. On increasing energies to 
$E=1\pm 0.1, 1.3\pm 0.1$, and $1.9\pm 0.1$ meV, we see excitations at 
${\bf Q}_{\rm IC}=(0,\pm(K+0.57),0)$ with $K=0,1$ and no scattering 
at ${\bf Q}_{\rm FM}$ [Figs. 2(b,c,d)].

Figure 2(e) shows the energy dependence of excitations along the
$[0,K,0]$ direction, which reveals two dispersionless excitations
at ${\bf Q}_{\rm IC}=(0,K+0.57,0)$ with $K=0,1$ and a dispersive excitation stemming
from the $(0,2,0)$ nuclear Bragg peak. The dispersionless excitations at
incommensurate wave vectors ${\bf Q}_{\rm IC}$ starting from $E=0.2$ meV must be spin fluctuations since low-energy acoustic phonons must be dispersive and 
originate from nuclear Bragg peak positions. To test if the dispersive excitation from
$(0,2,0)$ is indeed an acoustic phonon mode, we plot in Fig. 2(f) the 
2D image of excitations
along the $[0,K,2]$ direction. While the ${\bf Q}_{\rm IC}$ excitations
are no longer present, one can see a similar dispersive mode stemming from 
the nuclear Bragg peak $(0,2,2)$. By comparing the scattering 
intensity of these modes with nuclear structure factors, we conclude that 
the dispersive mode is the longitudinal acoustic phonon mode with a sound velocity of
$\sim$1000 m/s comparable to the sound velocity of UTe
measured by Brillouin light scattering \cite{Mendik1993}.
Figures 2(g) and 2(h) show constant-energy cuts along the $[0,K,0]$ direction
at different energies marked in Fig. 2(e).
At all energies probed, we see dispersionless incommensurate 
spin excitations centered around ${\bf Q}_{\rm IC}=(0,0.57\pm 0.02,0)$
with the dynamic spin correlation length of $\sim$12 \AA. This is close 
to the commensurate AF wave vector of $(0,0.5,0)$, thus indicating 
that spin fluctuations in UTe$_2$ are predominately AF in nature.

To see how incommensurate spin fluctuations
around ${\bf Q}_{\rm IC}$ change across $T_c$ and determine if there are strong FM spin fluctuations, we carried out  
energy scans at wave vectors ${\bf Q}_1$-${\bf Q}_6$ as marked in Fig. 2(a)
at temperatures $T=0.3, 2,$ and 12 K.
Figures 3(a) and 3(b) summarize the key results at 
the incommensurate wave vectors ${\bf Q}_1=(0,0.57,0)$ 
and ${\bf Q}_2=(0,1.57,0)$, respectively. To accurately determine 
the nuclear incoherent scattering backgrounds from the UTe$_2$ sample 
and the Cu sample holder, we chose 
${\bf Q}_3=(0,1.1,2)$ [Fig. 3(c)] and ${\bf Q}_4=(0,2,1)$ [Fig. 3(d)] since 
these positions are near the incommensurate and FM positions, respectively, 
but are sufficiently away from the nuclear Bragg peak positions. For possible FM spin fluctuations, we consider nuclear Bragg
peak positions ${\bf Q}_5=(0,1,1)$ [Fig. 3(e)] and 
${\bf Q}_6=(0,2,0)$ [Fig. 3(f)].

\begin{figure}[t]
\includegraphics[width=0.95\linewidth]{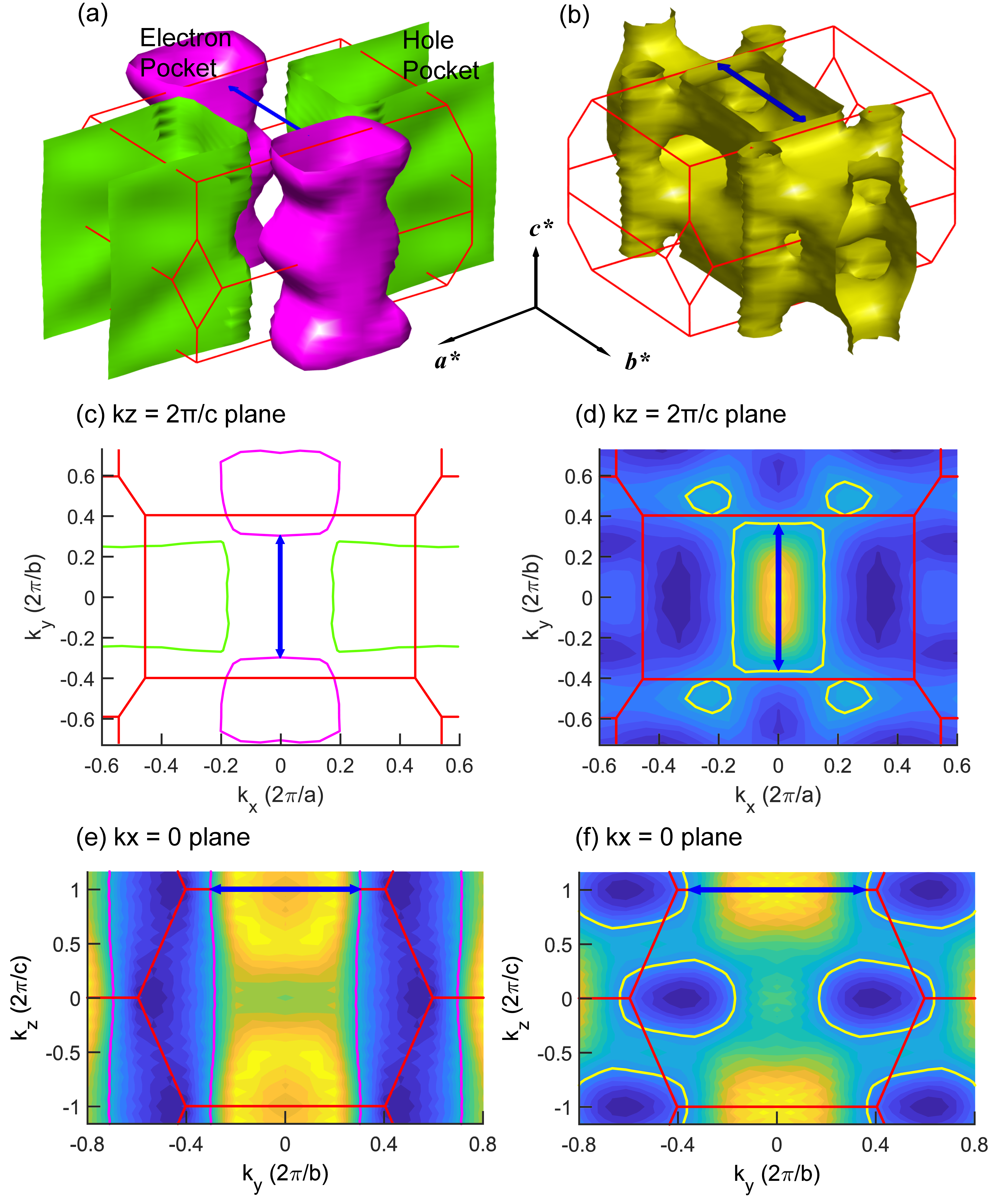}
\caption{(a,b) The 3D Fermi surfaces calculated by using DFT and   
treating U-$5f$ electrons 
as open core states. (b) The 3D Fermi surfaces of U-$5f$ electrons from DFT+DMFT calculations. Blue arrows on the top edge of the first BZ as well as those in subplots (c-f) indicate the spanning wave vector through Fermi surface nesting. (c-f) 2D cuts of the band calculation at the $k_z=2\pi/c$ plane and $k_x=0$ plane, respectively. In (c) to (f) the BZ is indicated by red lines, and the Fermi level of each band is marked with curves using the same color as used in the 3D plot in (a) and (b).}
\end{figure}

Figures 3(c) and 3(d) show the nuclear incoherent scattering background
at ${\bf Q}_3$ and ${\bf Q}_4$, respectively. As expected, the scattering is weakly
wave vector and temperature dependent between 0.3 K and 12 K. 
The green dashed lines in Figs. 3(a) and 3(b) are measured 
incoherent background scattering. The incommensurate spin fluctuations are clearly above the background scattering and follow the Bose population factor on warming from 0.3 K to 12 K. Figures 3(g) and 3(h) show the temperature
dependence of the imaginary part of the dynamic 
susceptibility $\chi^{\prime\prime}(E)$ at ${\bf Q}_1$ and ${\bf Q}_2$, obtained
by subtracting the incoherent scattering backgrounds and correcting for the Bose population
factor \cite{Dai2015}.  $\chi^{\prime\prime}(E)$ at both incommensurate 
wave vectors increase with increasing energy, but show no dramatic  
temperature dependence on warming from 0.3 K to 2 K across $T_c$, 
and to 12 K.  This is reminiscent of the
 temperature dependent $\chi^{\prime\prime}(E)$ in Sr$_2$RuO$_4$ \cite{Braden2004,Kunkem2017,Iida2020}, but clearly different from spin-singlet unconventional
heavy Fermion superconductors such as CeCoIn$_5$ \cite{Stock2008,YSong2020}, CeCu$_2$Si$_2$ \cite{Stockert2011}, and etc., where there is a strong enhancement of $\chi^{\prime\prime}(E)$, termed neutron spin resonance \cite{Eschrig2006,Wilson2006,GYu2009,Inosov2011}, in the pink marked energy region below $T_c$.  

Figures 3(e) and 3(f) summarize our attempt to extract FM spin fluctuations
in UTe$_2$, where the green dashed lines are incoherent 
scattering backgrounds measured at ${\bf Q}_3$ and ${\bf Q}_4$, respectively.
Compared with Figs. 3(a) and 3(b) at ${\bf Q}_{\rm IC}$, we see that 
FM spin fluctuations, if present, are much smaller
in magnitude and essentially vanish above $\sim$0.7 meV within the errors of
our measurements. Although temperature dependence of the scattering suggests
the presence of FM spin fluctuations, they do not dominate the spin fluctuations
spectra and neutron polarization analysis \cite{Steffens2019} may 
be necessary to conclusively identify FM spin fluctuations in UTe$_2$. 

To understand these results, we have also performed the 
electronic structure calculations of UTe$_2$ 
using the DFT+DMFT method \cite{AGeorges:1996,GKotliar:2006,KHaule:2010,RTutchton:2020}. 
In a heavy fermion metal such as UTe$_2$, there are two potential origins for the wave vector of incommensurate spin fluctuations. One is the RKKY interaction 
between the $5f$ moments, as appearing in a Kondo lattice 
model, which is 
determined by the electronic structure of the $spd$ conduction electrons. 
For this electronic structure, U-$5f$ electrons  are treated
as open core states (similarly via DFT to ThTe$_2$ \cite{wray2020}),  the calculated Fermi surface is shown 
in Figs. 4(a,c,e) \cite{SI}. Noticeably, the electron momentum transfer across
the two purple Fermi pockets is about $0.61 {\bf b}^*$, close to ${\bf Q}_{\rm IC}$ observed in the INS experiments. 
Another potential source for the wave vector of incommensurate spin fluctuations is the Fermi-surface 
nesting of the U-$5f$ heavy bands, which we have determined 
using the DFT+DMFT method \cite{SI}. The Fermi surface of the heavy bands
is shown in Figs. 4(b,d,f). Specifically, at $k_{z}=2\pi/c$,  the Fermi surface exhibits a rectangular shape,
and the electron momentum transfer across the short edges of the rectangular shape is about $0.72 {\bf b}^*$, which is slightly away from the observed ${\bf Q}_{\rm IC}$.
Therefore, the RKKY interaction of the $5f$ moments
is likely driving the incommensurate spin fluctuations of UTe$_2$, although 
the nesting of the strongly renormalized $f$-electron bands at the Fermi energy cannot
be ruled out.

In summary, we have discovered that the dominant spin fluctuations in UTe$_2$ are incommensurate near AF wave vector and  
extend to at least 2.6 meV. These results are consistent with DFT+DMFT calculations,
indicating that incommensurate spin fluctuations in UTe$_2$ may arise 
from the ${\bf Q}$-dependence of the RKKY interaction between the U-$5f$ moments.
 We expect these incommensurate spin fluctuations to play an important role 
in the development of the unconventional superconductivity.

The INS work at Rice is supported by the U.S. DOE, BES 
under grant no. DE-SC0012311 (P.D.). Part of
the material characterization efforts at Rice is supported by the Robert A. Welch
Foundation Grant Nos. C-1839 (P.D.). Research at UC San Diego was supported by the
U.S. DOE, BES under grant no. DEFG02-04-ER46105 (single crystal growth) and U.S. NSF under Grant No. DMR-1810310 (characterization of
physical properties). Work at Los Alamos was carried out under the auspices of the U.S. DOE National Nuclear Security Administration under Contract No. 89233218CNA000001, and was supported by the LANL LDRD Program.
The theory work at Rice has been supported by the U.S. DOE, BES 
under grant no. DE-SC0018197 and the Robert A. Welch Foundation Grant No. C-1411.
A portion of this research
used resources at the Spallation Neutron Source, a DOE Office of Science User
Facility operated by ORNL.

\end{document}